\documentclass[nologo,11pt,a4paper]{ETHpaper}
\usepackage{graphicx, epsfig, amsmath, amssymb,color}
\usepackage[square,numbers,sort&compress]{natbib}

\begin{document}

\newcommand{\mean}[1]{\left\langle #1 \right\rangle}
\newcommand{\abs}[1]{\left| #1 \right|}
\newcommand{\ul}[1]{\underline{#1}} \renewcommand{\epsilon}{\varepsilon}
\newcommand{\eps}{\varepsilon}
\renewcommand*{\=}{{\kern0.1em=\kern0.1em}}
\renewcommand*{\-}{{\kern0.1em-\kern0.1em}}
\newcommand*{\+}{{\kern0.1em+\kern0.1em}}

\renewcommand{\thefootnote}{ \fnsymbol{footnote} }

\title{How can social herding enhance cooperation?}

\titlealternative{How can social herding enhance cooperation?}

\author{Frank Schweitzer\footnote{Corresponding author:
    \url{fschweitzer@ethz.ch}}, Pavlin Mavrodiev, Claudio J. Tessone}

\authoralternative{F. Schweitzer, Pavlin Mavrodiev, C. J. Tessone}

\address{Chair of Systems Design, ETH Zurich, Weinbergstrasse 58, 8092 Zurich,
  Switzerland}

\reference{
Submitted  (2012). 
}

\www{\url{http://www.sg.ethz.ch}}

\makeframing
\maketitle

\renewcommand{\thefootnote}{\fnsymbol{footnote}}

\begin{abstract}
We study a system in which $N$ agents have to decide between two strategies $\theta_{i}$ ($i \in 1\ldots N$), for defection or cooperation, when  interacting with other $n$ agents (either spatial neighbors or randomly chosen ones).
After each round, they update their strategy responding nonlinearly to two different information sources: (i) the payoff $a_{i}(\theta_{i}, f_{i})$ received from the strategic interaction with their $n$ counterparts, (ii) the fraction $f_{i}$ of cooperators in this interaction.
For the latter response, we assume social herding, i.e.~agents 
adopt their strategy based on the frequencies of the different strategies in their neighborhood, \emph{without} taking into account the consequences of this decision.
 We note that $f_{i}$ already determines the payoff, so there is \emph{no additional} information assumed. A parameter $\zeta$ defines to what level agents take the two different information sources into account.
 For the strategic interaction, we assume a Prisoner's Dilemma game, i.e.~one in which defection is the evolutionary stable strategy. However, if the additional dimension of social herding is taken into account, we find instead 
a stable outcome where cooperators are the majority.
By means of agent-based computer simulations and analytical investigations, we evaluate the critical conditions for this transition towards cooperation. We find that, in addition to a high degree of social herding, there has to be a \emph{nonlinear} response to the fraction of cooperators.
We argue that the transition to cooperation in our model is based on \emph{less} information, i.e.~on agents which are not informed about the payoff matrix, and therefore rely on just observing the strategy of others, to adopt it. By designing the right mechanisms to respond to this information, the transition to cooperation can be remarkably enhanced.

\emph{Keywords: Prisoner's dilemma, social influence, mechanism design, nonlinear voter model}
\end{abstract}

\date{\today}

\section{Introduction}
\label{sec:Introduction}

\emph{Cooperation} is an abundant phenomenon in biological and social systems, but in most game-theoretical approaches \emph{defection} should be the rational strategy to choose. In order to solve this paradox, a vast number of literature has proposed modifications to the classical approach. They can be categorized along different directions:
\begin{itemize}
\item \emph{changes of the payoff structure:} lowering the costs of cooperation to make it more attractive in the first place is another form of "buying cooperation",
\item \emph{extension of the time horizon:} considering either repeated interaction, a memory
  for the strategy of the couterparts, calculating payoffs over a longer time interval, anticipating the future response to the
  own action,
\item \emph{considering spatial interaction:} the threshold for the outbreak of cooperation is lowered if agents' interaction is constrained to their nearest or second-nearest neighbors (as opposed to randomly chosen agents), or if agents can migrate between different spatial domains
\end{itemize}
We note that, particularly for biological systems, other additional mechanisms  have been considered \citep{nowak}, such as altruism, the role of kinship relations, selection mechanisms on the group level, etc.

In this paper, we add a new element to the discussion: social herding, i.e.~a mechanism that does \emph{not} take strategic considerations into account. Agents can observe the actions of others \emph{without} knowing their consequence. In a game-theoretical setting this means they cannot adopt a certain strategy based on payoff considerations because the payoff structure is not known to them. Thus, agents are just left with knowing the frequency of strategies either globally or in their neighborhood, and they choose their own strategy only based on the information about the frequency of these strategies. In our model, we assume that any agent can consider both the payoff-related  and the frequency-related information and weight their influence by a parameter $\zeta$, which is assumed to be the level of social herding. Precisely,
$\zeta\to 0$ results in purely payoff-driven decisions, $\zeta\to 1$ in pure social herding.

The case where social herding is dominant has been widely studied in binary opinion dynamics models \citep{schweitzer00b, fs-physa-02, Tessone2004, Castellano2009} where opinions are not necessarily related to payoff but rather to social norms. Thus, agents may adopt the opinion of a majority in order to minimize social conflicts, but they may not have a utility-based preference for either of these opinions. Instead their opinion results from a frequency-dependent decision. The so called \emph{linear voter model}, where the probability to choose a particular opinion is directly proportional 
to its frequency is a very common example for this. It is known to result in consensus, i.e.~the existence of only one opinion, asymptotically, but the outcome which opinion will dominate is not determined.
In the mean-field limit, this model always results in consensus of either of the two opinions. Starting e.g.~with the frequency $f_{1}$ of opinions $\theta=1$ and $f_0=1-f_1$ with $\theta=0$,
the probability that the final consensus state is $\theta_i = 1$ for all $i$ is $f_1$ \citep{holley75}.
Hence, a simple majority rule of social herding, as
expressed in the linear voter model, may not improve the situation for cooperation. Therefore, we turn to the class of \emph{nonlinear voter models} \citep{fs-voter-03} in Sect.~2. As we will also show analytically in Sect.~3, a nonlinear social herding by itself will not lead to a transition towards cooperation. Instead, it is needed the right level of social herding in combination with the right nonlinearity, to enhance cooperation.

What do we gain from such insights? First of all, a better understanding
of the fact that more information does not necessarily lead to a better
outcome (in this case, to cooperation). Common wisdom would suggest that
it is always better to have more information, e.g.~to choose among more
alternatives, to determine their consequences in advance, and thus to
\emph{reduce the risk} associated with making the wrong decision. What
seems to be an optimal strategy on the individual level, turns out to
lead to the lock-in into unfavorable situations on the global level.
For example, in experiments on the \emph{wisdom of crowd} effect, it was
shown that more information about the guesses of other agents, combined
with social influence, leads to a failure in the predictions \citep{lorenz2011social}. Also, in a network formation model of agents sharing knowledge it was shown that \emph{best response}, 
i.e.~the choice of partners based on knowing all alternatives, resulted in a worse global performance as compared to a situation where just the
next best partner was accepted \citep{koenig-book-09}.
As we point out with this work, to leave the trap of defection also crucially depends on using less of the available information, or to have a considerable fraction of less informed agents.

Second, from our insights we can derive mechanisms to improve the outcome in systems of strategically interacting agents. Mechanism design can be seen as the engineering part of economics. It allows to propose rules, or algorithms,  for interactions that avoid the system getting trapped in suboptimal states. Some of these algorithms, such as the nowadays famous ``Gale-Shapley" algorithm \citep{gale1962college}, are basically related to combinatorial optimization problems. I.e., they propose a solution \emph{for} the agents \emph{without} involving the agents in finding it, themselves. Systems design, the way we see it, aims instead at proposing new ways of \emph{interaction} at the agent level, in order to arrive at more favorable solutions at the system's level. Our paper gives a lucid example of this kind of systems design, by proposing a different way of combining information an individual agent already has. This still leaves room for the forces of self-organization to act, but restricts the possible negative outcome.

\section{Basic Model}
\label{sec:Model}

\subsection{Combining social herding and strategic interaction}
\label{sec:interaction}

We consider a system with $N$ agents. Each agent $i \in 1 \ldots N$ is
characterized by two individual variables which may change over time:
$\theta_{i}(t)$ shall describe the agent's strategic behavior when
interacting with other agents, whereas $\zeta_{i}(t)$ shall describe how
much the agent is prone to social influence. We adopt the definition of
\textit{social influence} as the psychological tendency of individuals to adhere
to and behave according to the expectations of its local
neighborhood \citep{Kahan1997}. In this sense, our approach belongs to
a wider class of models which do not restrict herding behavior to
perfectly rational agents \citep{Parker2005}.

In an economic context,
$\theta_{i}$ refers to the strategy of a utility maximizing agent,
chosen from a (discrete) set $\sigma$ of possible strategies.  We use the
standard game theoretical setting of a Prisoner's Dilemma (PD) game,
i.e. $\sigma \in \{0,1\}$, where the strategic behavior $\sigma=0$ refers
to \emph{defection} $(D)$ and $\sigma=1$ to \emph{cooperation} $(C)$.

We assume that each agent plays a 2-person (non-iterated) game with $n$
other agents which are located in its neighborhood. The completion of these $n$ games is called a round. From
each of these interactions the agent receives a payoff which depends both
on the strategic behavior of the agent itself and on the opponents'. The game structure describing a single interaction between two agents can be summarized by the standard payoff matrix of a 2-person game:
\begin{equation*}
  \label{eq:pd}
  \begin{array}[c]{c|c|c|}
    & \theta_{j}=1 & \theta_{j}=0 \\ \hline
    \theta_{i}=1 & R/R & S/T \\
    \theta_{i}=0 & T/S & P/P
  \end{array}
\end{equation*}
Suppose, agent $i$ has chosen to cooperate, then its payoff is $R$ if the
other agent $j$ has also chosen to cooperate (without knowing about the
decision of agent $i$), but $S$ if agent $j$ defects. On the other hand,
if agent $i$ has chosen to defect, then it will receive the 
payoff $T$ if agent $j$ cooperates, while it will receive $P$ if agent $j$ defects.

In this paper, we will restrict the discussion to the PD game, but we
note that our investigations can be extended to other games that result
from different values of $R$, $S$, $T$ and $P$ \citep{fs-lb-acs-02}. For the particular case of the PD game, the payoffs have to fulfill the following two inequalities:
\begin{equation}
  \label{pd-ineq1}
  T > R > P > S \;; \quad 2\,R > S+T
\end{equation}
The known standard values are $T=5$, $R=3$, $P=1$, $S=0$. This implies that, in a
so-called one-shot game (no repeated interaction), defection
$\sigma=0$, is the rational strategy because it rewards the higher payoff
for an agent $i$ no matter whether the opponent chooses $C$ or $D$. As
this argument applies to both agents, one can expect that on the system
level a global defective behavior emerges. Because of this, the PD game has become a paradigmatic model to study different mechanisms of
transition towards a global cooperative behavior \citep{axelrod2006,Szabo2007a}, a question that has puzzled the
scientific community for decades.

Let us define the degree of cooperation on the system's level by the
total number of cooperating agents, $N_{1}(t)$ relative to the total
population $N$.  Since the number of agents is constant, the global frequencies
$f_{\sigma}$ of cooperating and defecting agents are given by
\begin{eqnarray}
  \label{nconst}
  N&=&\sum_{\sigma}N_{\sigma}=N_{0}+N_{1}= \mathrm{const.} \;;
  \quad \sigma
  \in \{0,1\}, \nonumber \\
  f_{\sigma}&=&\frac{N_{\sigma}}{N} \;; \quad f \equiv f_{1} = 1-f_{0}.
\end{eqnarray}
In the following, the variable $f$ shall refer to the global frequency of
cooperators.

The interaction of each agent with $n$ other agents in a 2-person game
results in ${N \choose n}$ different possibilities to choose a
partner. As the result of these interactions that may occur
\emph{independently}, but \emph{simultaneously} \citep{Hauert2001,fs-lb-acs-02},
agent $i$ receives a total payoff $A_{i}(\theta_{i})$ which depends both
on its own strategy $\theta_{i}$ and the strategies of the $n$ different
partners. Let us assume that $n_{0}$ of these partners have chosen to
defect, whereas $n_{1}=n-n_{0}$ partners have chosen to cooperate. Then
the total payoff from these $n$ interactions reads:
\begin{equation}
  \label{payoff-i}
  A_{i}(\theta_{i})= \delta_{1,\theta_{i}}\, \Big[n_{1}\,R+n_{0}\,
  S\Big]   +  \delta_{0,\theta_{i}} \,\Big[n_{1}\,T+n_{0}\,P\Big],
\end{equation}
where $\delta_{x,y}$ means the Kronecker delta, which is 1 only for $x=y$
and zero otherwise. Dividing by $n$ gives the scaled total payoff:
\begin{equation}
  \label{payoff-in}
  a_{i}(\theta_{i},f_{i})= \frac{A_{i}(\theta_{i})}{n}=
\delta_{1,\theta_{i}}\, \Big[f_{i}\,R+(1-f_{i})\,
  S\Big]   +  \delta_{0,\theta_{i}} \,\Big[f_{i}\,T+(1-f_{i})\,P\Big],
\end{equation}
where $f_{i}=n_{1}/n=1-n_{0}/n$ gives the fraction of cooperating agents
agent $i$ interacts with. Assuming e.g. that agent $i$ interacts with its
neigbors, $f_{i}$ gives the \emph{local} frequency of cooperators. If on
the other hand agent $i$ interacts with $n$ randomly chosen agents, the
probability to choose a cooperator is directly proportional to the global
fraction $f$. I.e. in the so-call \emph{mean-field approach} we set
$f_{i}\equiv f$.

Strategic considerations imply that agent $i$ pays attention to the
scaled payoff $a_{i}(\theta_{i},f_{i})$ expected from the interaction
with $f_{i}$ cooperators, which of course also depends on its own
strategy $\theta_{i}$. A nonlinear function $\mathcal{G}(a_{i})$ shall
consider the way agent $i$ combines the information about the different
payoffs $a_{i}(\theta_{i},f_{i})$ and $a_{i}(1-\theta_{i},f_{i})$
resulting from its possible strategic choice. This shall used below to define the transition rate for an agent to change between strategies, therefore we conveniently normalize  $\mathcal{G}(a_{i})$ to one. In a very general way, we
assume:
\begin{equation}
\mathcal{G}(a_{i}) = \frac{\exp{[\beta_{i}\ a_{i}(\theta_{i},f_{i})]}}{
\exp{[\beta_{i}\ a_{i}(\theta_{i},f_{i})]}+\exp{[\beta_{i}\ a_{i}(1\-\theta_{i},f_{i})]}}.
\label{eq:g-ai}
\end{equation}
Eq.~(\ref{eq:g-ai})  has the form of a logit-function well established
in decision theory \citep{Blume1993,Thurstone1994,McKelvey95}.
The parameter $\beta_{i}$ allows agents to individually weight differences between the
payoffs. $\beta_{i}\to 0$ represents the limit of random choice
between strategies, $\mathcal{G}(a_{i}) \to 1/2$, whereas
$\beta_{i}\to \infty$ means that even small differences in payoff lead
to an immediate switch between $\mathcal{G}(a_{i})=0$ and
$\mathcal{G}(a_{i})=1$.
For small values of $\beta_i$, the  $\mathcal{G}(a_{i})$ tends to one if
the expected payoff times the  $a_{i}(\theta_{i},f_{i})$ from stategy
$\theta_{i}$ is much larger than the expected payoff
$a_{i}(1-\theta_{i},f_{i})$ from the opposite strategy $1-\theta_{i}$.
and it tends to zero in the opposite case. If both payoffs become
comparable, $G(a_{i})$ is about $1/2$. Intermediate values of $\beta_{i}$ allow for a
smooth transition between the two strategic cases.

We note that for sufficiently small values of $\beta_{i}$ Eq. (\ref{eq:g-ai}) can be
approximated by the linear function
\begin{equation}
  \label{eq:gai-lin}
  \mathcal{G}(a_{i}) \approx \frac{1}{2}\left[1 +
  \frac{\beta_{i}}{2}\big\{a_{i}(\theta_{i},f_{i}) - a_{i}(1\-\theta_{i},f_{i}) \big\}\right],
\end{equation}
i.e.~agents pay attention to the \emph{difference} between the two possible
payoffs.

The situation becomes different if the agent is unable to calculate
the expected payoff. In our model, we assume that the agent then
rather pays attention to the action of the majority and tends to
imitate this without knowing about the consequences. Thus, agent $i$
only responds to the information associated with the frequency which
shall be described by a logit-function similar to Eq.~(\ref{eq:g-ai}):
\begin{equation}
  \label{eq:ffi}
  \mathcal{F}(f_{\theta_{i}})=\frac{\exp{[2\beta_{i}
    \kappa_i (f_{\theta_{i}})\; f_{\theta_{i}} \- 1]}}{
\exp{[2\beta_{i} \kappa_i(f_{\theta_{i}})\; f_{\theta_{i}}\-1]}+
\exp{(-[2\beta_{i} \kappa_i(f_{\theta_{i}}) \; f_{\theta_{i}}\-1])}}.
\end{equation}
$f_{\theta_{i}}$ describes the local frequency of agents playing
strategy $\theta_{i}$ in the neighborhood of agent $i$, and
$f_{1-\theta_{i}}=1\-f_{\theta_{i}}$ is the local frequency of agents
playing the opposite strategy. Both frequencies being equal,
$\mathcal{F}(f_{\theta_{i}})=\mathcal{F}(f_{1-\theta_{i}})=1/2$.
Again, for sufficiently small $\beta_i$, from a linear approximation in Eq.~(\ref{eq:ffi}) we find,
\begin{equation}
  \label{eq:ffi-lin}
  \mathcal{F}(f_{\theta_{i}})  \approx
  \beta_{i} \kappa_i(f_{\theta_{i}})\, f_{\theta_{i}}.
\end{equation}
$\kappa_i(f_{\theta_{i}})$ is a nonlinear response function to
consider a weighted influence of the frequency \citep{fs-voter-03} as we will investigate
below.  $\kappa_i(f_{\theta_{i}})$ may also depend on the time an agent has kept its current strategy, or opinion \citep{tessone2008a, tessone2008b}.
We emphasize that for the so-called linear voter model, 
$\kappa_i(f_{\theta_{i}})$ is simply a
constant $\kappa$ that does not depend on the frequency. So $\beta_{i}
\kappa$ can be scaled to one, which means that for the linear voter
model we simply arrive at
$\mathcal{F}(f_{\theta_{i}})=f_{\theta_{i}}$. Thus, the response of
agent $i$ is directly proportional to the local frequency of agents
playing strategy $\theta_{i}$.

After having defined the agent's response to strategic information and
to social herding, we use the individual parameter $\zeta_{i}$ to
weight these two different influences. Specifically, we define the
transition rate for agent $i$ to switch from strategy
$(1\-\theta_{i})$ to the opposite strategy $\theta_{i}$ as follows:
\begin{equation}
  \label{eq:w}
w(\theta_i|(1\-\theta_i),f_{i},\zeta_{i})= (1-\zeta_{i})\, \mathcal{G}(a_{i})
+ \zeta_{i}
\, \mathcal{F}(f_{\theta_{i}}).
\end{equation}
For $\zeta_{i}\to 0$, we cover the limit case of strategic interaction
in PD game, for $\zeta_{i}\to 1$, we arrive at the limit case of pure
social herding, i.e. imitation behavior without calculating the resulting
consequences.

\subsection{Specifying the transition rates}
\label{sec:spec}

Before describing the system's dynamics by means of a master equation in the following section, it will be handy to write down the
transition rates of Eq.~(\ref{eq:w}) more specifically. The transition rates apply for a \emph{frequency dependent} process,
i.e. they do not depend on the specific sequence of interaction. In this paper, we fix the number of independent, but simultaneous
2-person games to $n=4$, which is convenient to compare random interactions with local ones on a regular lattice. Hence, the
relevant frequencies have only discrete values $f_{i}\equiv k_{i}/n$ where $k_{i}\equiv n_{1}=0,1,2,3,4$ is the actual number of
cooperating agents, agent $i$ is interacting with. 
On the other hand, random interactions can be approximated by the so-called mean field approximation, where $f_{i}=f$, the global fraction of cooperators.

Dropping the individual index $i$ for the moment, we have to distinguish between two different transition rates,
$c_{k}(\zeta)=w(1|0,k,\zeta)$, i.e. the transition from defection to cooperation dependent on $k$ cooperating agents, and
$d_{k}(\zeta)=w(0|1,k,\zeta)$, i.e. the transition from cooperation to defection under the same conditions. Both of these rates are
comprised of two parts, one resulting from strategic behavior ($\tilde{c}_{k}$, $\tilde{d}_{k}$), the other one resulting from
social herding ($\hat{c}_{k}$, $\hat{d}_{k}$),
\begin{equation}
  \label{eq:ckdk}
  c_{k}(\zeta)= (1-\zeta) \tilde{c}_{k} + \zeta \hat{c}_{k}\;; \quad
d_{k}(\zeta)= (1-\zeta) \tilde{d}_{k} + \zeta \hat{d}_{k}.
\end{equation}
For the terms ($\hat{c}_{k}$, $\hat{d}_{k}$) related to social herding, we use the linear approximation, Eq. (\ref{eq:ffi-lin}), i.e. for the specified neighborhood $n=4$,
\begin{equation}
  \label{eq:lin-cd}
\hat{c}_{k}=\frac{k}{4}\beta\kappa_k  \;; \quad
\hat{d}_{k}=1-\frac{n-k}{4}\beta\kappa_k.
\end{equation}
Again, for the linear voter model with $\kappa_{k}\equiv \kappa$ and
the ($\hat{c}_{k}$, $\hat{d}_{k}$) would simply result from the set of values $\{$0, 1/4, 2/4, 3/4, 1$\}$. In order to use nonlinearities in the frequency response, we rather prefer to specify the ($\hat{c}_{k}$, $\hat{d}_{k}$) by discrete values $\alpha_{0}$, $\alpha_{1}$, $\alpha_{2}$ as shown in Table (\ref{a1a2})
  \begin{equation}
  \begin{array}[c]{c|cc|cc}
    \; f=k/n \quad & \quad \tilde{c}_{k} \quad & \quad \tilde{d}_{k} \quad &  \hat{c}_{k} & \hat{d}_{k} \\ \hline
    0  & \tilde{c}_{0} & \tilde{d}_{0} & \alpha_{0} & 1\- \alpha_{0}\\
    1/4 & \tilde{c}_{1} & \tilde{d}_{1}& \alpha_{1} & 1\- \alpha_{1}\\
    2/4& \tilde{c}_{2} & \tilde{d}_{2} & \alpha_{2} & \alpha_{2}\\
    3/4& \tilde{c}_{3} & \tilde{d}_{3} & 1\- \alpha_{1} & \alpha_{1}\\
    1 & \tilde{c}_{4} & \tilde{d}_{4} & 1\- \alpha_{0} & \alpha_{0}\\
  \end{array}
  \label{a1a2}
\end{equation}
The parameter $\alpha_{0}$ describes the transition of a cooperator (defector) towards defection (cooperation) if surrounded by cooperators (defectors) solely based on \emph{social herding}. Because agents with such strategies to follow are absent in the neighborhood, $\alpha_{0}$ should be consequently zero, even if there is a strong \emph{strategic} incentive for a cooperator to switch towards defection if surrounded by cooperators. Hence, considering only social herding, pure cooperation and pure defection are ``absorbing" states for the dynamics of the system.
This can be avoided by choosing $\alpha_{0}=\epsilon$, a very small value that allows for occasional random changes of the strategies \citep{fs-voter-03}, but in this paper we choose $\alpha_{0}=0$.

\begin{figure}[htbp]
  \includegraphics[width=7.cm]{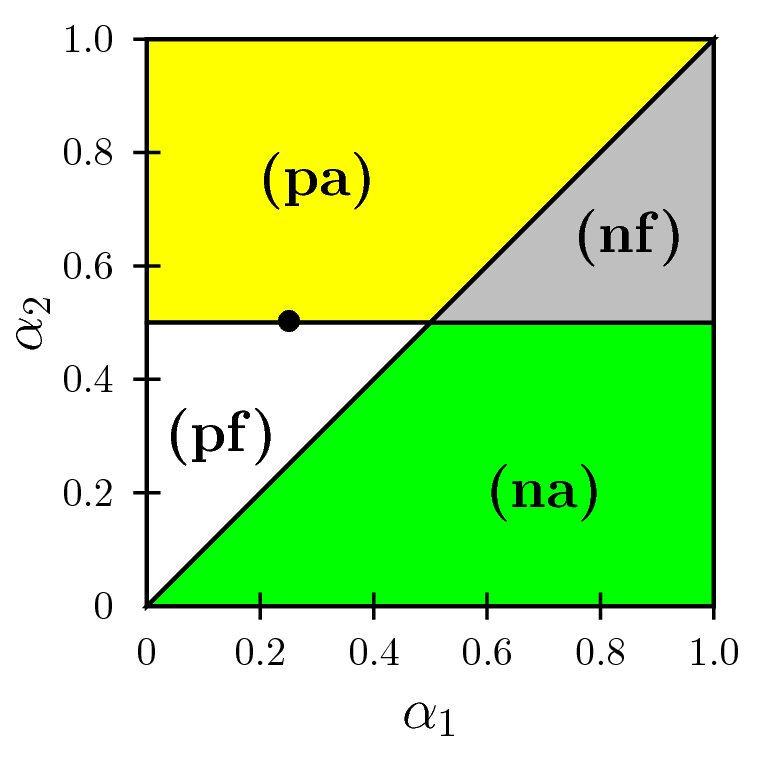}   \hfill
  \centering
    \includegraphics[width=7.cm]{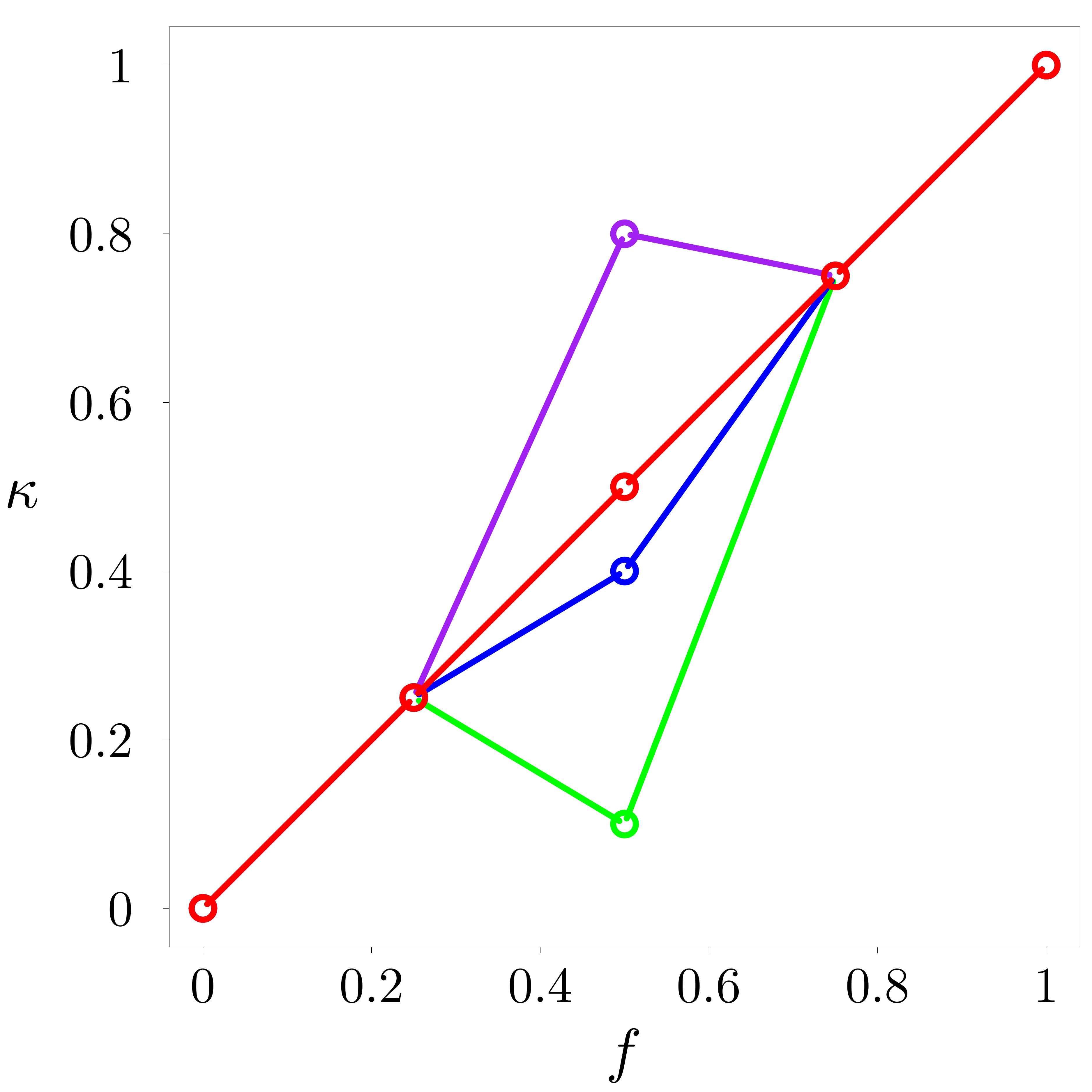}
  \caption{(left) Parameter space $(\alpha_{1},\alpha_{2})$ to define the nonlinearity in social herding (see also Table \ref{a1a2}). The different regions are explained in the text. We use the (pa) region, defined by Eq. (\ref{eq:allee}).  (right) Linear voter model (red line) and deviations controlled by $\alpha_{2}$ at $f=0.5$. }
  \label{fig:non}
\end{figure}

Possible combinations of $(\alpha_{1},\alpha_{2})$ define a parameter space to distinguish between different forms of social
herding, as shown in Fig. \ref{fig:non} (left). Positive frequency dependence (pf) means that the probability to change to the
opposite strategy monotonously increases with the frequency of that strategy in the neighborhood, also known as ``majority
voting''. Negative frequency dependence (nf) means the opposite, i.e. the probability monotonously decreases with the frequency,
also known as ``minority voting''. On the other hand, (pa) and (na) define parameter regions with non-monotonous dependence. For example, (pa) means
an increase of the probability as long as the opposite stragety is not the majority, also known as voting against the trend, while
(na) describes constellations with a strong amplification of minority strategies. We note that the so-called ``voter point'' that
represents the the linear voter model --where $\alpha_{1}=1/4$ and $\alpha_{2}=2\alpha_{1}=1/2$ are strictly proportional to $k$--
is on the border between the (pf) and (pa) parameter regions.  For our investigations, we will consider a scenario where the
nonlinearity is only represented by $\alpha_{2}$, whereas $\alpha_{1}$ is chosen according to the linear voter model. Four
possible cases which refer to the (pf), (pa), (na) and the linear voter
model are shown in Fig. \ref{fig:non} (right)

It remains to specify the payoff related terms ($\tilde{c}_{k}$, $\tilde{d}_{k}$) which follow directly from
Eq. (\ref{eq:g-ai}). Here, we assume the deterministic limit $\beta_{i}\to 0$, for which we get
$\mathcal{G}(a_{i})=\Theta\big[a_{i}(\theta_{i},f_{i}) - a_{i}(1\-\theta_{i},f_{i}) \big]$, where $\Theta[y]$ is the Heavyside
function, which is one if $y>0$ and zero otherwise. I.e. $\mathcal{G}(a_{i})$ is either one or zero dependent on whether the
payoff for the changed strategy is larger or less than the payoff resulting from the current strategy.
Taking into account the payoff relations, Eq. (\ref{pd-ineq1}), we verify that the expected payoffs, Eq. (\ref{payoff-in}), for defectors, $a(0,f)$, are \emph{always} larger than the corresponding ones for cooperators, $a(1,f)$, regardless of the fraction of cooperators in the neighborhood. I.e. in non-repeated games as considered here, defection is an evolutionary stable strategy. Hence, in the deterministic limit of stategic interaction, we have always $\tilde{c}_{k}=0$ and $\tilde{d}_{k}=1$. This can be rightly assumed as the worst-case scenario because, considering only  a strategic point of view, the system will always end up in pure defection. The most important thing is to identify conditions where an additional social herding allows not only to avoid this trap, but also to let the dynamics to converge to pure cooperation.

The observant reader may have noticed that we have interpreted  $\beta_{i}$ differently for social herding (where we assumed that it is just small) and for strategic interaction (where $\beta \to 0$ was assumed). This is not a contradiction. In fact, $\beta$ quantifies the randomness in following the different information, and we can assume that the payoff related attention is much higher and less prone to errors than the response to the behavior of neighbors. In general, we may distinguish between $\tilde{\beta}_{i}$ and $\hat{\beta}_{i}$ for the different responses, but this is not applied here.

\subsection{Dynamics to change the strategy}
\label{sec:strategy}

In the previous section, we have defined the ``rules'' for agents to change their strategy dependent on both strategic information and social herding. Most agent-based models, at this point, would continue with extensive computer simulations to probe the parameter space for some non-trivial results. We will certainly follow with computer simulations as well, however we are also interested in some analytical insights into the model which would allow us to predict the system's dynamics without testing every possible parameter combination. For this reason, we need to specify the dynamics of agents in a more formal way, on two different levels, (a) on the micro level of the individual agent, and (b) on the macro level, describing the fraction of cooperators in the system.

For the micro level, we use a stochastic approach, i.e.~we deal with the probability $p_{i}(\theta_{i},t)$ that agent $i$ uses strategy $\theta_{i}$ at time $t$. As explained before, this probability depends on the strategies of agents in the neighborhood of agent $i$ expressed by the vector $\ul{\theta}_{i}= \{\theta_{i_1},\theta_{i_2},\ldots,\theta_{i_{n}}\}$. Hence, $p_{i}(\theta_{i},t)$ is defined as the marginal distribution:
\begin{equation}
  \label{eq:marginal}
  p_{i}(\theta_{i},t)=\sum_{\ul{\theta}_{i}^{\prime}} p(\theta_{i},\ul{\theta}_{i}^{\prime},t).
\end{equation}
The summation is over all possible distributions $\ul{\theta}_{i}^{\prime}$.  Specific realizations of these distributions shall
be denoted as $\ul{\sigma}$. For $n=4$, there are $2^{n}$ possible realizations.  For the time-dependent change of $
p_{i}(\theta_{i},t)$ we assume the following master equation:
\begin{equation}
  \label{master}
  \frac{d}{dt}p_i(\theta_i,t)=\sum_{\ul{\theta}_{i}^{\prime}} \Big[
  w(\theta_i|(1\-\theta_i),\ul{\theta}_{i}^{\prime})\;
  p(1\-\theta_i,\ul{\theta}_{i}^{\prime},t) - w(1\-\theta_i|\theta_i,\ul{\theta}_{i}^{\prime})\;
  p(\theta_i,\ul{\theta}_{i}^{\prime},t)\Big].
\end{equation}
This equation considers all possible processes that may lead to an increase or decrease in the probability that agent $i$ uses
strategy $\theta_{i}$ given the neigborhood distribution $\ul{\theta}_{i}$, with the transition rates
$w(\theta_i|(1\-\theta_i),\ul{\theta}_{i}^{\prime})$, $w(1\-\theta_i|\theta_i,\ul{\theta}_{i})$. Note that these are not the
transition rates defined in Eq. (\ref{eq:w}), which only depend on the local frequency $f_{i}$, but not on the neighborhood
distribution $\ul{\theta}_{i}$. In order to map the two, we have to consider how many specific realizations of the distribution
$\ul{\theta}_{i}$ may lead to the same $f_{i}$. Taking the example $\ul{\sigma}=\{0010\}$, there are exactly ${4 \choose 1}$
different possibilities to realize $f_{i}=1/n$. Hence, transforming the master Eq.~(\ref{master}) that depends on the neighborhood
distribution $\ul{\theta}_{i}$ into one that only contains the respective local frequency $f_{i}$ results in a combinatorial
prefactor of ${n \choose k}$. Using again the specific notations $c_{k}$, $d_{k}$, Eq. (\ref{eq:ckdk}) for the transition rates,
we can rewrite the master equation (\ref{master}) now as
\begin{equation}
  \label{eq:mastercomplete}
  \frac{d}{dt}p_i(1,\zeta,t)=\sum_{k=0}^{n}
  {n \choose k} \Big[ c_{k}(\zeta) \; p(0, k/n,\zeta,t)
  - d_{k}(\zeta) \; p(1, k/n,\zeta, t) \Big].
\end{equation}
The corresponding master equation for $p_{i}(0,\zeta,t)=1-p_{i}(1,\zeta,t)$ follows likewise.
Note that in Eq. (\ref{eq:mastercomplete}) we have chosen the individual parameter $\zeta_{i}$ to be a constant $\zeta$. I.e. whereas the local frequency $f_{i}=k/n$ changes over time because of concurrent decisions of neighboring agents about their strategies, $\zeta$ is, in this paper, assumed to be a global control parameter the impact of which will be discussed together with the computer simulations.

With this, we have a bottom-up description of the system's dynamics given by $N$ stochastic equations, Eq. (\ref{eq:mastercomplete}), which are coupled because of the overlapping neighborhoods of agents, expressed in terms of $f_{i}$. On the other hand, on the macroscopic level we have to deal with the probability $P(f,\zeta,t)$ to find a given fraction of cooperators, $f$, at time $t$, assuming the social herding factor $\zeta$. The dynamics can again be specified by a stochastic equation:
\begin{equation}
  \label{master-P}
  \frac{d}{dt}P(f,\zeta,t)=\sum_{f^{\prime}} \Big[
  W(f|f^{\prime},\zeta)\;
  P(f^{\prime},\zeta,t) - W(f^{\prime}|f,\zeta)\;
  P(f,\zeta,t)\Big].
\end{equation}
$f^{\prime}$ denotes all possible deviations from a given value $f$ that can be reached during one time step by means of the transition rates $W(f^{\prime}|f,\zeta)$. These are not identical with the individual transition rates, Eq.~(\ref{eq:w}), but aggregated rates that take into account all possible ways to change $f$. The smallest change of $f\equiv N_{1}/N$, Eq. (\ref{nconst}), is the addition or substraction of a single cooperator, i.e. $f^{\prime}\in \{(N_{1}+1)/N;\;(N_{1}-1)/N\}$. The individual equivalent for such processes is given by Eq.~(\ref{eq:ckdk}), where the terms $c_{k}(\zeta)$ describe the transition of a single defector into a cooperator, and the $d_{k}(\zeta)$ the opposite transition. Hence, we find for the aggregated transition rates
\begin{eqnarray}
\label{Wplusminus}
W(f+1/N|f,\zeta) &\equiv& W_{+}(f,\zeta) = \sum_{k=0}^{n}{n \choose k}f^{k}(1-f)^{n-k}\; c_{k}(\zeta) \nonumber \\
W(f-1/N|f,\zeta) &\equiv& W_{-}(f,\zeta) = \sum_{k=0}^{n}{n \choose k}f^{n-k}(1-f)^{k}\; d_{n-k}(\zeta).
\end{eqnarray}
The combinatorial prefactors preceeding the $c_{k}(\zeta)$ and $d_{k}(\zeta)$ result from the various ways to choose agents with
$n=4$ neighbors, $k$ of which could be cooperators given the gobal fraction of cooperators $f$. Here, we have used the so-called mean-field assumption that replaces the frequencies $f_{i}$ of the individual neigborhoods by the global value $f$. With the specific values for $c_{k}(\zeta)$ and $d_{k}(\zeta)$ given by Eqs.~(\ref{eq:ckdk}) and (\ref{a1a2}), the dynamics on the systemic level is also completely specified. In the following, we will use the dynamics on the micro level for carrying out computer simulations, while the dynamics on the macro level will be used for analytical investigations.


\section{Results of Computer Simulations}
\label{sec:comp}

\begin{figure}[htbp]
   \centering
   \includegraphics[width=7.cm]{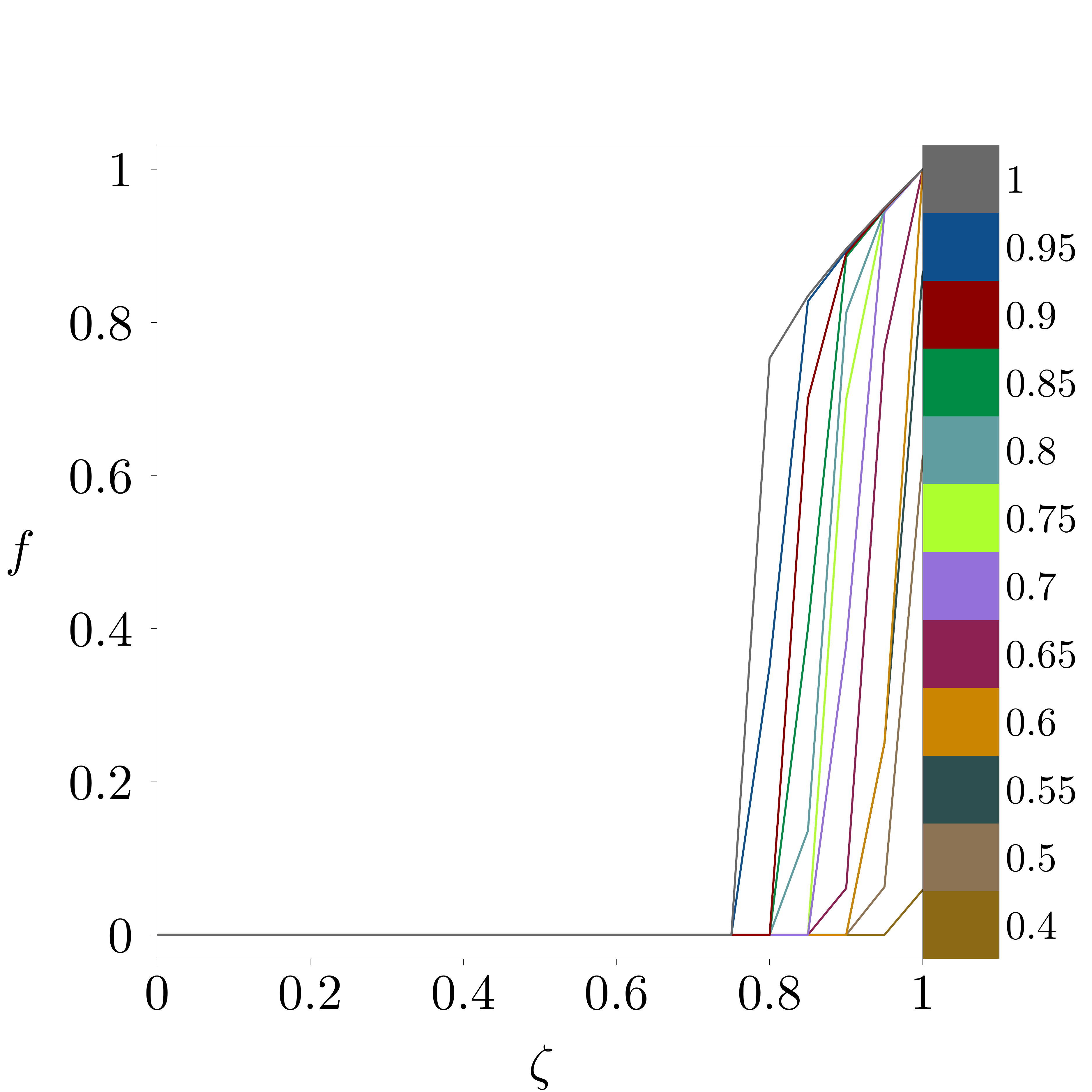}
   \caption{Global fraction of cooperation $f$ dependent on the level of social herding $\zeta$. $\alpha_{1}= 0.25$ is fixed, $\alpha_{2}$ varies between 0.4 and 1.0 according to the color scale. System size $N$=400.}
   \label{fig:zeta}
 \end{figure}

We now use the dynamics specified in Eq. (\ref{eq:mastercomplete}) to run agent-based computer simulations for different sets of parameters. According to Eqs. (\ref{eq:ckdk}), (\ref{a1a2}), we only need to vary the weight $0 \leq \zeta \leq 1$  and the parameters $0\leq (\alpha_{1},\alpha_{2})\leq 1$ assigned to  the social herding of the agents. Regarding their strategic decision, everything is already defined, and with $\tilde{c}_{k}=0$, $\tilde{d}_{k}=1$ defection remains the only choice.
This ``worst case scenario'' can be only changed because of a considerable amount of social herding, in which agents copy the strategy of their neighbors regardless of the payoff assigned to it. This is shown in Fig.~\ref{fig:zeta}. Below a critical level for social herding, $\zeta\approx 0.7$, only defection remains. For $\zeta>0.7$ we observe different levels of cooperation which depend on the combination of $\zeta$ and $\alpha_{2}$. If $\zeta>0.8$, cooperation even becomes the majority, i.e.~$f>0.5$, but only for large values of $\zeta$ and $\alpha_{2}$ full cooperation, $f\to 1$, is reached. This issue is further investigated below.

\begin{figure}[htbp]
   \centering
   \includegraphics[width=7.cm]{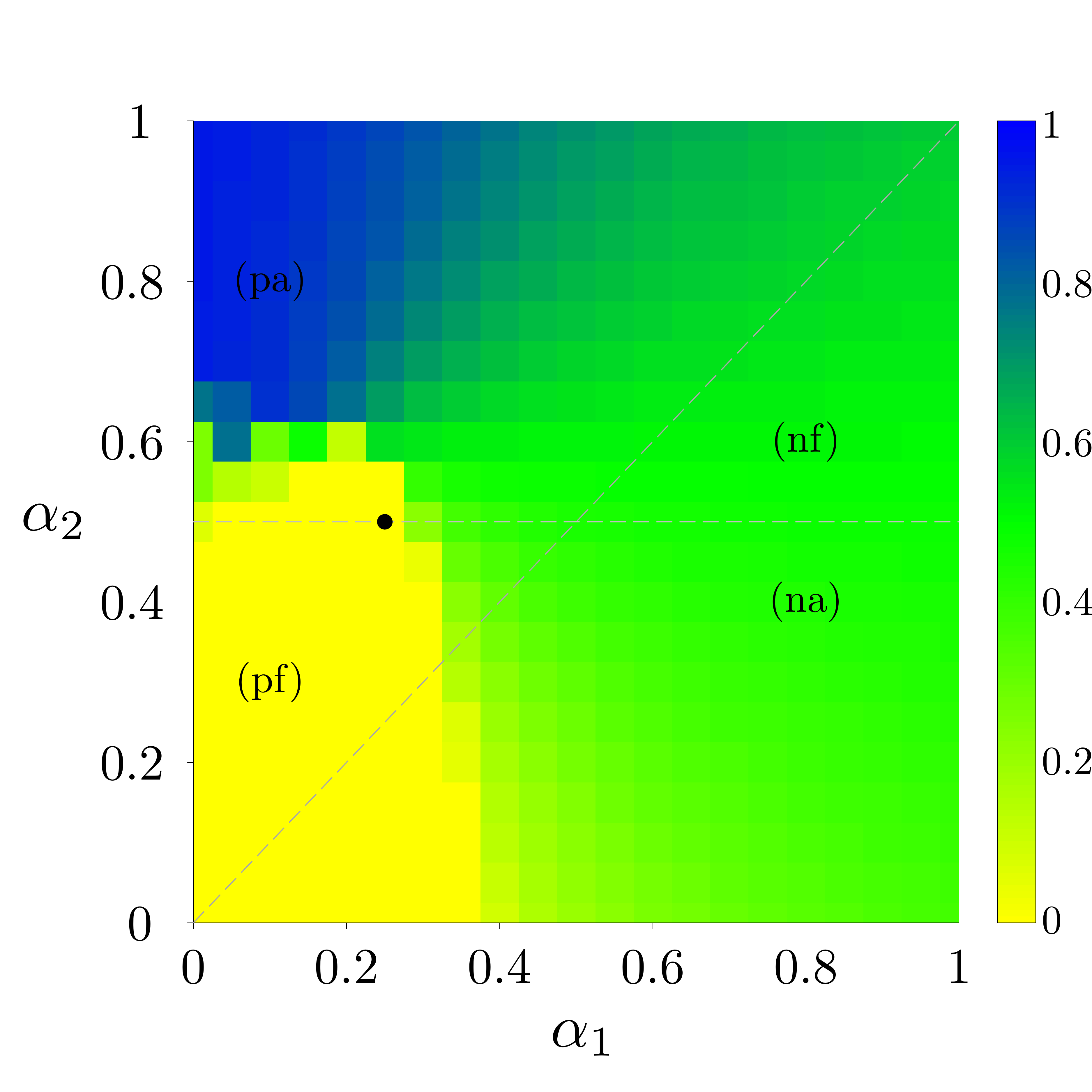}
   \caption{Fraction of cooperation (color scale) dependent on the nonlinearities in social herding, defined by $\alpha_{1}$, $\alpha_{2}$. Fixed level of social herding $\zeta$=0.95. The four different areas are defined in Fig. \ref{fig:non}(left). $\bullet$ indicates the linear voter model. Szstem size $N$=400.}
   \label{fig:a1a2-color}
 \end{figure}

The role of the nonlinearity in social herding, expressed in terms of $\alpha_{1}$, $\alpha_{2}$, is further investigated in Fig. \ref{fig:a1a2-color}, given a supercritical level of social herding. We see that there is an \emph{optimal nonlinearity} to enhance cooperation, i.e. $\alpha_{1}$, $\alpha_{2}$ have to be chosen such that they belong to the area of ``positive allee'' (pa) effects. This area is defined by the inequalities (see also Eq.~(\ref{a1a2})),
\begin{equation}
  \label{eq:allee}
  0\leq \alpha_{1} \leq \alpha_{2};\;  (1-\alpha_{1})\leq \alpha_{2} \leq 1.
\end{equation}
It describes a response where the transition toward a given strategy \emph{increases} with the frequency of that strategy as long as that strategy is \emph{not} the majority, i.e. minority strategies are favored. A special case where $\alpha_{1}$ is taken from the linear voter model, whereas $\alpha_{2}$ is larger than 0.5 is shown in Fig. \ref{fig:non} (right).
We note in particular that social herding according to the \emph{linear} voter model will \emph{not} allow the transition toward cooperation, which will be further substantiated by analytical results in the next section.
Further, all forms of  the transition rates that \emph{monotonously} increase with the frequency, indicated by the (pf) area, will \emph{not} lead to cooperation. Social herding in this case only amplifies defection.

 Assuming the right choice of parameters for the transition to cooperation, we can now take a look how the dynamics evolve in
 space. We have chosen a two-dimensional regular lattice with Von-Neumann neighborhood, where each agent interacts with $n=4$ local neighbors. Initially, we assume a small
 cluster of cooperating agents. \emph{Without} social herding, this cluster would immediately disappear in the next time step because all
 agents will choose defection, which is the rational choice to maximize their payoff. We observe instead a spreading of cooperation, i.e. an invasion of the cooperating strategy into the domain of defectors. The cooperating agents, however, do not form compact clusters. A minority fraction of defectors will always survive and their spatial distribution in small clusters across the domain of cooperators continues to change in time. I.e. we never reach a stationary state in space, despite that the global fraction of both strategies, on average, reaches an equilibrium.

We further note that there is a critical size for the initial cluster of cooperators to grow. This has been already discussed in detail for pure  PD games on a regular lattice \citep{Hauert2001,fs-lb-acs-02}, and in opinion dynamics models \citep{Tessone2004}.  Now, the addition of supercritical social herding of course reduces these requirements.
Is it worth mentioning that, starting from random initial conditions in a spatially extended system, we find that a vanishingly small initial density of cooperators is enough to trigger the final state. The reason for this stems from the fact that, if the system is large enough, one cluster of cooperators larger than the critical size will appear by chance. This cluster will be sufficient to trigger the outbreak of cooperation.
Here, however, we will not dig further on this discussion. Instead,
the initial conditions and parameter constellations for the \emph{outbreak of cooperation} will be further discussed for the mean-field case, in the next section.
 \begin{figure}[htbp]
   \centering
   \includegraphics[width=2.5cm]{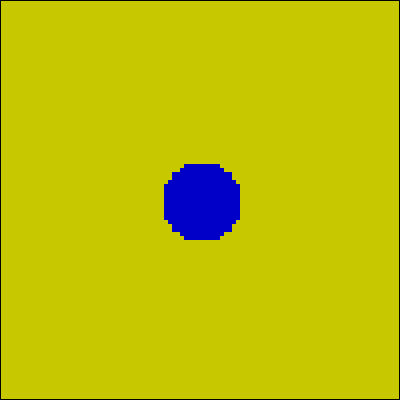}
   \includegraphics[width=2.5cm]{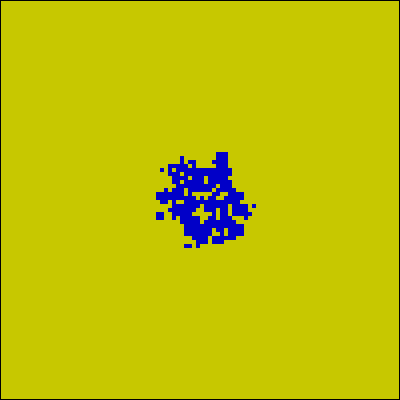}
   \includegraphics[width=2.5cm]{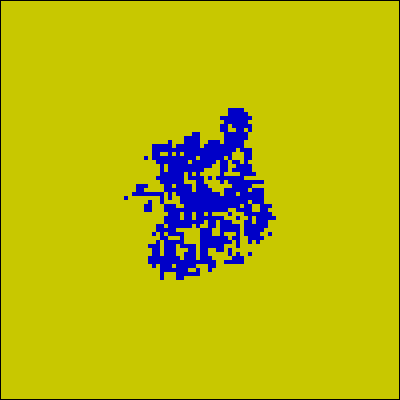}
   \includegraphics[width=2.5cm]{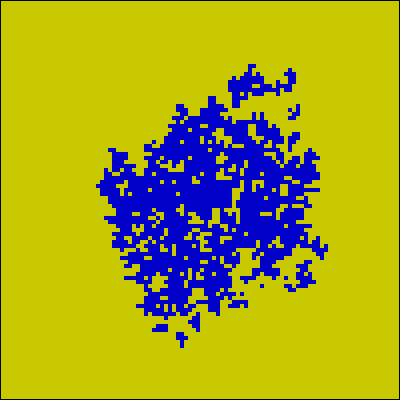}
   \includegraphics[width=2.5cm]{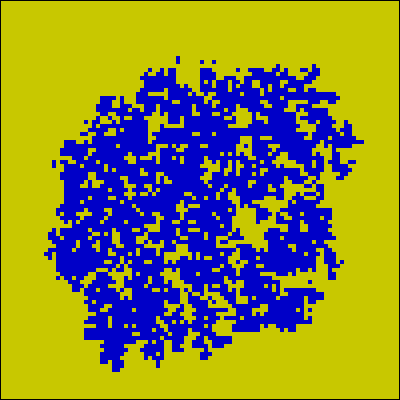}
   \includegraphics[width=2.5cm]{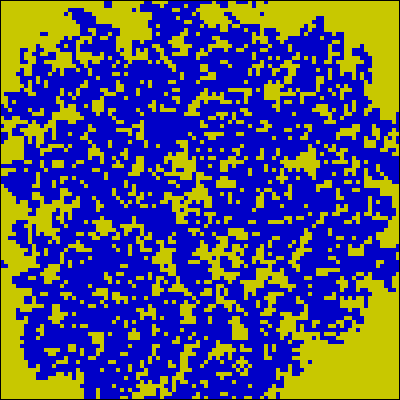}
   \caption{Snapshots of the transition toward cooperation at times $t$=0, 10, 20, 50, 150, 500. $N$=$10^4$ agents are placed on a regular lattice and interact each with their $n$=4 spatial neigbhors. Dark color (blue) indicates cooperators, light color (yellow) defectors. Parameters $\alpha_1=0.25$, $\alpha_2=0.7$, $\zeta = 0.95$.
System is a two-dimensional regular lattice with Von-Neumann neighborhood and size $N$=400.
}
   \label{fig:snapshots}
 \end{figure}

\section{Mean-field investigations}
\label{sec:mean}

\subsection{Calculating the effort}
\label{sec:effort}

We verified by means of computer simulations that there is indeed a way of utilizing social herding to boost cooperation. Now, we try to illustrate this finding by some analytical considerations. As a first step, we want to calculate the ``effort'' to transfer the system into a majority of cooperators. Considering only the strategic dimension, this effort should be very high because there is a strong incentive to defect. On the other hand, social herding may help in this situation because it neglects the payoff differences. So, it is particularly important in the first stage of the phase transition.

A formal approach to calculate the effort starts from the master equation (\ref{master-P}) on the systemic level, in the mean-field limit. The detailed balance condition, which is a specific form of the equilibrium condition $dP(f,t)/dt =0$, requires that the net probability fluxes are balanced, i.e.
\begin{equation}
  \label{eq:balance}
  W(f|f-1/N,\zeta)\;
  P^{0}(f-1/N,\zeta) = W(f-1/N|f,\zeta)\;
  P^{0}(f,\zeta),
\end{equation}
where $P^{0}(f,\zeta)$ denotes the equilibrium probability distribution which is independent of $t$.
This equation is recursive and, using $f=N_{1}/N$, Eq. (\ref{nconst}), can be re-formulated as:
\begin{equation}
P^{0}(f,\zeta) = P^{0}(0,\zeta)\;\prod_{i=1}^{N_{1}}\frac{W\left(\left. \frac{i}{N}\right|
\frac{i-1}{N},\zeta\right)}{W\left(\left. \frac{i-1}{N}\right|\frac{i}{N},\zeta\right)}.
\label{null}
\end{equation}
The normalization $P^{0}(0,\zeta)$ can be found by enforcing $\sum_{i=0}^{N} P^{0}(i/N,\zeta)=1$ and the transition rates are given by
Eq. (\ref{Wplusminus}). We visualize the equilibrium probability distribution by means of a potential $\Omega(f,\zeta)$ that has its minimum where $P^{0}(f,\zeta)$ has its maximum, i.e. it represents the ``effort'' of reaching a given equilibrium state,
\begin{equation}
  \label{eq:potential}
  P^{0}(f,\zeta)= \exp\{-\Omega(f,\zeta)\},
\end{equation}
where $\Omega$ is given by
\begin{equation}
\Omega(f,\zeta) = -\ln P^{0}(0,\zeta)-
\sum_{i=1}^{N_{1}}\ln \left[
\frac{W\left(\left. \frac{i}{N}\right|
\frac{i-1}{N},\zeta\right)}{W\left(\left. \frac{i-1}{N}\right|\frac{i}{N},\zeta\right)}
\right]
\label{eq:omega}
\end{equation}
Figure \ref{fig:zeta} shows the effort $\Omega(f,\zeta)$ as a function of
the global fraction of cooperators $f$ and the level of social herding
$\zeta$, which acts as a control parameter. We observe that for very low
values of $\zeta$ the effort is a monotonously increasing function of
the frequency $f$. Given a fraction of cooperators, $f=0.2$, and small
$\zeta$, it becomes more and more difficult, or unlikely, to find
a larger fraction of cooperators (red line). Considering instead a high
level of social herding, e.g  $\zeta$ about 0.85. there is a \emph{monotonous decrease} of the
effort with an increasing fraction of cooperators. I.e. starting from a
supercritical level of social herding, the outbreak and the increase of
cooperation becomes very likely (green line).
\begin{figure}[htbp]
  \centering
  \includegraphics[width=10.cm]{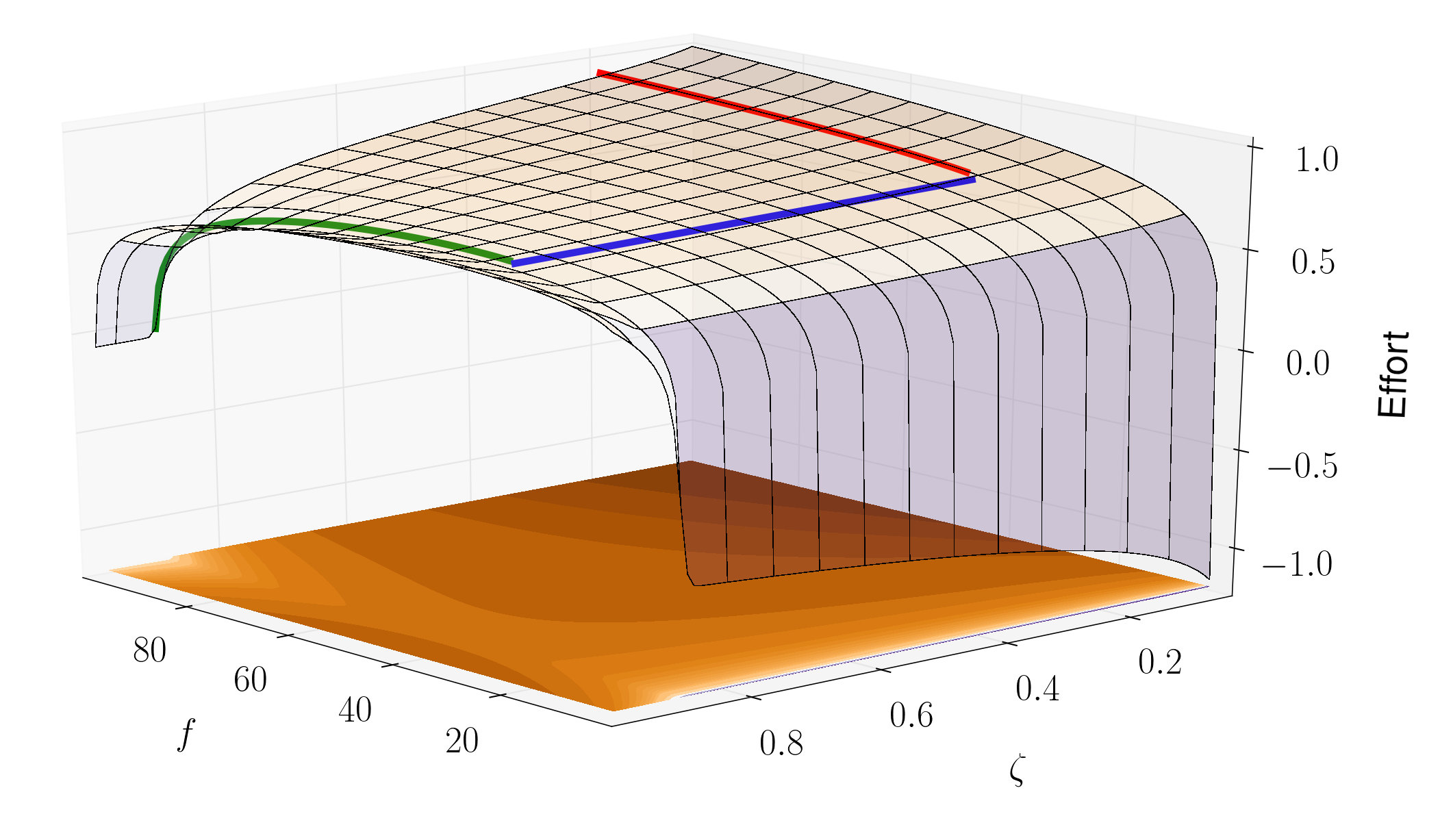}
  \caption{Effort  $\Omega(f,\zeta)$, Eq. (\ref{eq:omega}) dependent on
the global fraction of cooperators $f$ and the level of social herding $\zeta$. The nonlinearity is specified by $\alpha_{1}$=0.25, $\alpha_{2}$=0.85.}
  \label{fig:effort}
\end{figure}

The observant reader will notice in Figure \ref{fig:effort} for large $\zeta$ the \emph{nonmonotonous dependence} of the effort on the fraction of cooperators. I.e. there is a critical region around of $f\approx 0.2$ below which defection becomes the most probable state. This relates to the critical cluster size of cooperators in Fig. \ref{fig:snapshots} to allow the transition toward cooperation. However, there is a noticable difference underlying both results. Fig. \ref{fig:effort} is based on the mean-field limit, i.e. there is no spatial correlation between interacting agents, whereas Fig. \ref{fig:snapshots} assumes a spatial neigborhood defined by the regular lattice. In fact, it is known that spatial interaction enhances cooperation \citep{roca2009effect,Schweitzer2005,fs-lb-acs-02}. Already small, randomly formed clusters of cooperators are sufficient for the outbreak of cooperation, whereas random interaction results in a much larger threshold.

\subsection{Competition dynamics}
\label{sec:competition}

Eventually, we can also derive a deterministic dynamics for the global fraction of cooperators, $f(t)$, in the mean-field limit. Basically, there are two ways of deriving this. One starts from the stochastic dynamics on the microscopicl level, $p_{i}(\theta_{i},t)$, Eq. (\ref{master}) and is discussed in detail in \citep{fs-voter-03}. The other one starts from the stochastic dynamics on the macroscopic level, $P(f,\zeta,t)$ , Eq. (\ref{master-P}). The expected value for the global fraction of cooperators then follows from
\begin{equation}
  \label{eq:expected}
  \mean{f(\zeta,t)}= \sum_{f^{\prime}}P(f^{\prime},\zeta,t),
\end{equation}
where $f^{\prime}$ denote all possible realizations of $f$. Using the master equation (\ref{master-P}), we arrive at the deterministic dynamics
\begin{equation}
  \label{eq:det}
  \frac{d\mean{f(\zeta,t)}}{dt} = {W}_{+}(f,\zeta)\;  (1-\mean{f}) - {W}_{-}(f,\zeta)\; \mean{f},
\end{equation}
where the aggregated transition rates $W_{+}(f,\zeta)$, $W_{-}(f,\zeta)$ are given by Eq.~(\ref{Wplusminus}). Assuming a narrow probability distribution in equilibrium, $P^{0}(f,\zeta)$, the expected value  $\mean{f^{0}(\zeta)}$ can be approximated by the maxima of $P^{0}(f,\zeta)$. In particular, the deterministic dynamics will converge to those areas where $P^{0}(f,\zeta)$ is largest, or where $\Omega(f,\zeta)$ has its minima, shown in Fig. \ref{fig:effort}. While we do not argue about the specific global dynamics at intermediate times (which can be governed by stochastic influences in particular in early stages), we can see the late stage of the dynamics as a ``quasi-stationary'' motion along the valley in the potential landscape shown in Fig.~\ref{fig:effort}, provided $\zeta$ chosen large enough.

\begin{figure}[htbp]
  \centering
  \includegraphics[width=7.cm]{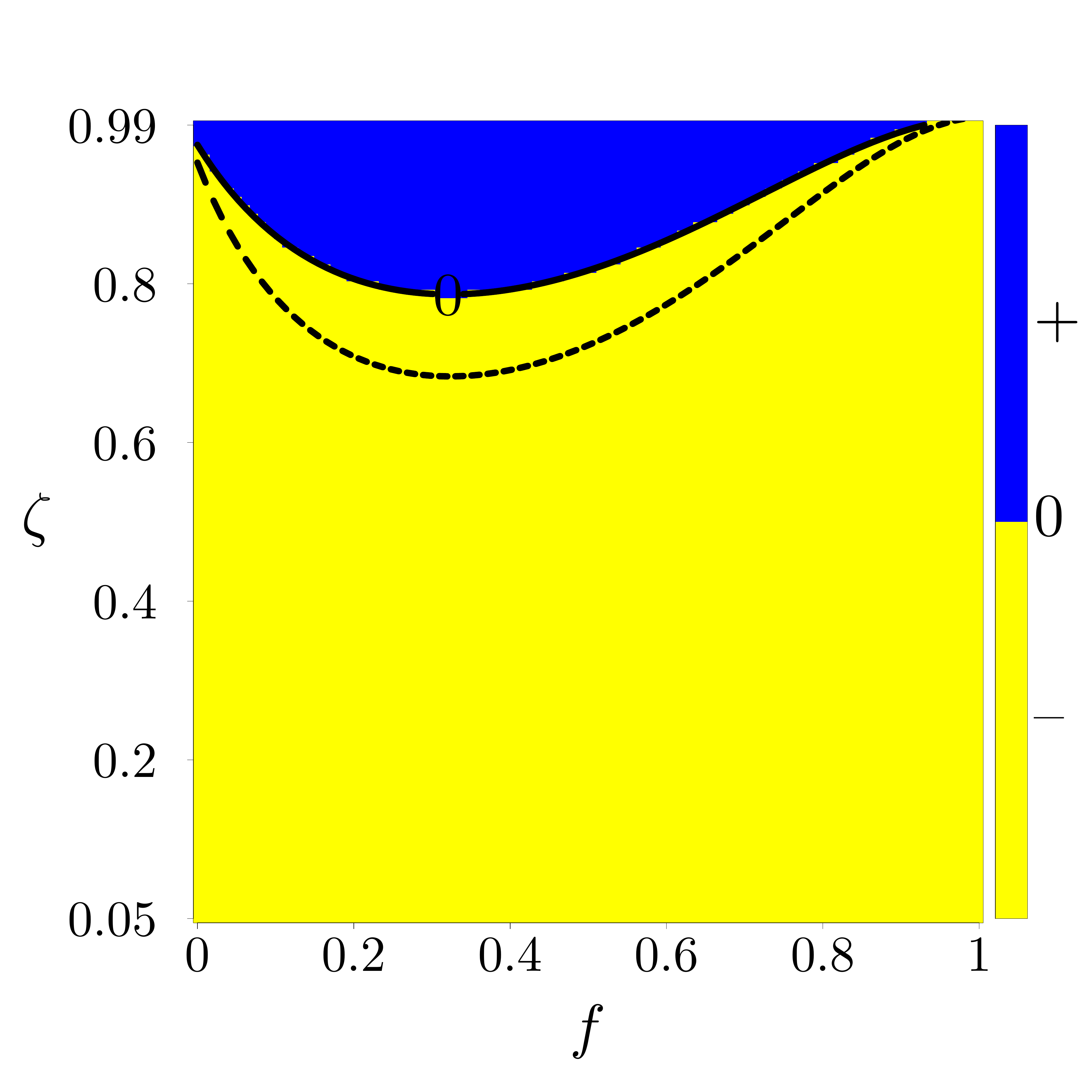}
  \caption{Difference of the fitness values $E_{1}(f,\zeta)-E_{0}(f,\zeta)$ dependent on the fraction of cooperators, $f \in [0.02,0.99]$, and the level of social herding,  $\zeta \in [0.05,0.99]$. Nonlinearity
parameters: $\alpha_{1}=0.25$, $\alpha_{2}=0.8$.}
  \label{fig:diff}
\end{figure}

We can rewrite Eq.~(\ref{eq:det}) which basically describes the ``replication'' of cooperators at the global scale, to make it more alike to the known replicator equation,
\begin{equation}
  \label{eq:replicator}
  \frac{d\mean{f(\zeta,t)}}{dt} =  \mean{f}\; (1-\mean{f})  \Big[ {E}_{1}(f,\zeta)- {E}_{0}(f,\zeta)\; \Big].
\end{equation}
The two terms $E_{1}$ and $E_{0}$ are the fitness values associated with the two different strategies.  The fraction of cooperation will grow if
the fitness of cooperation ${E}_{1}(f,\zeta)$ is larger than the fitness of defection ${E}_{0}(f,\zeta)$, which both depend on the global level of cooperation and the level of social herding,
\begin{equation}
  \label{eq:fitness}
  {E}_{1}(f,\zeta)= \frac{W_{+}(f,\zeta)}{f}\;; \quad  {E}_{0}(f,\zeta)= \frac{W_{-}(f,\zeta)}{1-f}.
\end{equation}
To evaluate the fitness values,
one should note the stricly nonlinear depencence of the transion rates on
$f$, (\ref{Wplusminus}). Fig. \ref{fig:diff} shows the difference
$E_{1}-E_{0}$ on the whole range of $f$ and $\zeta$. We emphasize that this graph holds for fixed values of the nonlinearity parameters $\alpha_{1}$, $\alpha_{2}$, i.e. it adds another dimension to  Fig. \ref{fig:a1a2-color}, which was obtained for a fixed herding level $\zeta$. Fig. \ref{fig:diff} also clearly shows the influence of the initial fraction of cooperators, $f(0)$, for the mean-field case. Assuming e.g. a fixed value of $\zeta$=0.85, we see that the fraction of cooperators $f(t)$ can be increased in time only if $f(0)$ is between 0.15 and and 0.6. While the lower bound has an  intuitive meaning as the minimum threshold to start cooperation, the upper bound is less obvious. It results indeed from the influence of the \emph{nonlinear} social herding, which does not simply support cooperation if that is the strategy of the majority. We recall that social herding does not assume any "value" related to the strategies. Hence, for the example considered, the maximum fraction of cooperators is given by
$f=0.6$. A higher level of social herding, or different values for the nonlinearities, may increase this fraction up to about one, i.e. full cooperation.

Another way of expressing the dynamics of Eq.~(\ref{eq:replicator}) is through
\begin{equation}
  \label{eq:select}
  \frac{d \mean{f(\zeta,t)}}{d t}= \mean{f(\zeta,y)} \big( E_{1} - \mean{E}\big)\;;\quad
  \mean{E}=\sum_{\sigma}E_{\sigma}\mean{f_{\sigma}}= E_{1}\mean{f} +E_{0}(1-\mean{f}).
\end{equation}
As long as $E_{1}$ is larger than the average fitness, $\mean{E}$, the fraction of cooperators in the system is able to grow, but
one has to recognize that, because of the time dependence of $\mean{f(t)}$ and its implicite feedback on $E_{\sigma}$, $\mean{E(t)}$
evolves over time as well. Hence, Eq. (\ref{eq:select}) describes a nonlinear selection process for each of the strategies dependent on the parameters describing strategic interaction and social herding.

For some special cases, we are able to derive closed form solutions of the competition dynamics expressed by Eqs. (\ref{eq:det})-(\ref{eq:select}). In the absense of any social herding, $\zeta\=0$,
we just have to count in the transition rates from strategic interaction, which are $\tilde{c}_{k}\=0$, $\tilde{d}_{k}\=1$. This results in $E_{1}(f,\zeta\=0)=0$ and $E_{0}(f,\zeta\=)=1$, i.e. the dynamics reads
$\mean{f(t)}=f(0)\exp\{-t\}$,
which means that cooperation dies out, exponentially. In the opposite case, $\zeta=1$, i.e. absense of any strategic interaction, Eq. (\ref{eq:replicator}) can be solved for the case of the linear voter model, which implies $\hat{c}_{k}=k/4$ and $\hat{d}_{k}=1\-(k/4)$, Eq. (\ref{eq:lin-cd}). We then find $E_{1}(f,\zeta\=1)=E_{0}(f,\zeta\=1)$, i.e. the fitness of both strategies, which are actually mere labels without any payoff assigned, is the same. This results in the dynamics $\mean{f(t)}=f(0)$, i.e. a \emph{conservation} of the initial fraction of cooperators, on average. This is known as one of the puzzles associated with the linear voter model, i.e. individual realizations of the dynamics, e.g. using  stochastic simulations, always lead to convergence with $f\to 0$ or $f\to 1$, but averaging over many runs reveals that the frequency at which cooperators or defectors dominate is equal to their initial fraction $f(0)$.

These two limiting cases allow us to position the dynamics if $0 < \zeta< 1$, i.e. the influence of both strategic interaction and social herding at the same time. For social herding, let us  first assume the case of the linear voter model as described above. We can then verify that the closed solution for the dynamics of cooperators is given as:
\begin{equation}
  \label{eq:closed}
  \mean{f(\zeta,t)}=f(0) \;\exp\{(\zeta\-1) t\}
\end{equation}
which is similar to the case of only strategic interaction, except that the time scale for the extinction of cooperators is stretched by the factor $(1\-\zeta)$. This is an important result because it demonstrates that \emph{linear} social herding will \emph{not} prevent the extinction of cooperation, not even for large $\zeta$. Hence, in order to turn defection into cooperation, we essentially need a \emph{high} level of \emph{nonlinear} social herding, i.e. the right $\zeta$ and $\alpha_{2}$ values.

Considering a nonlinearity where $\alpha_{1}=1/4$ but $\alpha_{2}\neq 2/4$, we find from Eq. (\ref{eq:det})
\begin{equation}
  \frac{d\mean{f(\zeta,t)}}{dt} = \mean{f} \left\{ \zeta \left[ 1 +
3 \mean{f}\left(1-\mean{f}\right)^2 \left(2 \alpha_2 -1\right)\right] - 1\right\}
\label{eq:fnon}
\end{equation}
For $\alpha_{2}=2/4$, the solution reduces to Eq. (\ref{eq:closed}), whereas for $\zeta=1$ we arrive at the mean-field equation for
the nonlinear voter model, only \citep{fs-voter-03}. In order to make cooperation, $\mean{f}=1$, a stable fixed point for the full dynamics, the following condition for $\alpha_{2}$ has to be met:
\begin{equation}
 \frac{1}{2} + \frac{1-\zeta}{6\zeta \mean{f}\left(1-\mean{f}\right)^2} < \alpha_2 \leq 1
\label{eq:alpha2}
\end{equation}
which implies $1/[1+3\mean{f}(1-\mean{f})^{2}]<\zeta<1$. This inequality
can be only met for a considerable high level of social herding. The feasible range of $(f,\zeta)$ values that is consistent with a given value of $\alpha_{2}$, e.g. $\alpha_{2}$=0.8, is shown in Fig. \ref{fig:diff}. The maximum range resulting from $\alpha_{2}=1$ is also shown in the same Figure by the dashed line. We note again that, even if Eq. (\ref{eq:alpha2}) is fulfilled, the dynamics does not necessarily converge to $f\to 1$. Dependent on the parameters $\{\zeta,\alpha_{1},\alpha_{2}\}$ also lower equilibrium fractions of cooperators may be reached, i.e. we find a \emph{coexistence of coooperation and defection.}


\section{Conclusions}
\label{sec:concl}

In this paper, we have explored a new route towards cooperation. This route differs from many other attempts, most of which are  rooted in traditional or evolutionary game theory, where the transition toward cooperation is induced by specific neighborhood relations, repeated interactions, discounted payoffs over long time horizons, indirect reciprocity, favorable strategy mutations, the enforcement of social norms, etc \citep{szolnoki2009impact,Szabo2007a,helbing2008migration,fs-lb-acs-12,tessone2012} All of these propositions either improve the payoff of the cooperating strategy or provide, in one or another way, \emph{additional information} agents may consider  when making a strategic decision.

Our approach is much simpler, by not changing payoffs at all, but only counting on the information agents alredy have if they simultaneously play  a $2$-Person PD game with their $n$ neighbors (which can be spatial neighbors, or randomly chosen). This information is the local fraction of cooperators, $f_{i}=n_{1}/n$, and defectors, $(1-f_{i})$, of an agent, that also enters the calculation of the payoff, Eq.~(\ref{payoff-in}). That means there is \emph{no} additional information assumed. We argue instead that agents, at the same time, respond to this information in two different ways, as summarized in Eq. (\ref{eq:w}).
In a strategic interaction, they choose the strategy $\theta_{i}$ that will lead to the highest payoff $a_{i}(\theta_{i},f_{i})$, whereas in the case of social herding they simply respond to the local frequency of each strategy in a nonlinear manner,  $\mathcal{F}(f_{\theta_{i}})$. In some sense, the second way assumes \emph{less} information because no payoff matrix needs to be known. This implies that both strategies are seen as equally valuable.

The parameter $\zeta_{i}$ gives a weight to these two different ways of utilizing the information associated with $f_{i}$. In Eq. (\ref{eq:w}) we have assumed $\zeta_{i}$ to be an individual parameter, which means that agents dependent on their internal preferences or access to knowledge (such as a known payoff) can give different weights to these two responses. In this paper, however, we did not further explore this source of heterogeneity, but kept it as a global parameter, constant and the same for all agents. This limit case is equivalent of assuming a population of agents, a fraction $\zeta$ of which \emph{only} follows social herding, whereas a fraction $(1-\zeta)$ \emph{only} considers strategic interactions. This allows to interpret our main result about a critical $\zeta$ to turn a population of defectors into cooperators in a more general manner: $\zeta$ can be seen as the minimal fraction of agents following only social herding, to enable the transition to cooperation. With respect to the access
to information, we can interpret this finding as follows: if the information about the payoff matrix is know to all agents, they will --in the given Prisoner's Dilemma setting-- collectively choose defection (which is the suboptimal state). However, if only a small fraction of agents (about 20\%) (see Fig. \ref{fig:zeta}) has information about the payoff matrix  and the majority will just respond to the decision of others by means of \emph{nonlinear} social herding, this can drive the system towards a state where  cooperation is the dominant strategy. To put it succinctly: \emph{less} information (or a larger fraction of uninformed agents) will  lead to \emph{more} cooperation.

This interesting and important conclusion still relies on choosing the right nonlinear social herding in response to the local (or global) fraction of cooperators. We have demonstrated that the \emph{linear} response, where the probability to choose a strategy is directly proportional to the fraction of that strategy in the neighborhood (or the population), \emph{fails} to enhance cooperation. Instead, we have to choose a nonlinearity, expressed in terms of the parameters $\alpha_{1}$, $\alpha_{2}$, from the region of \emph{positive allee} (pa) effects (Fig. \ref{fig:zeta}). As a minimal condition for the transition towards cooperation, all transition rates can be (but not necessarily have to be) chosen according to the linear voter model, except
$\alpha_{2}$, which has to be above the critical value $0.5$ to break the tie in case of an equal fraction of cooperators and defectors.
Further, the combination of $\zeta$ and $\alpha_{2}$ also determines the maximum level of cooperation that can be reached using the two different responses.

Our finding tells that social herding matters most in tie situations, which is also similar to another class of group decision models \cite{galam2008}.  To design a mechanism that influences social herding only in
this situation also provides a quite ``cost-efficient'' solution in that we will not need to enforce a decision against the
majority, to allow for the transition toward cooperation.
Agents can still follow the strategy of the majority --just in the
undecided case, we need to ensure that the symmetry is broken into the ``right'' direction.

Eventually, we wish to point out that in this paper we have discussed a
kind of worst-case scenario where, in the absense of social herding,
defection is the only stable state for the system. Even for this case,
our proposed mechanism excels in transferring defectors into cooperators,
on the population level. We can leverage other model ingredients to
further facilitate this transition. For example, we could count in
stochastic changes of the strategy as already considered in the strategic
component $\mathcal{G}(a_{i})$, Eq.~(\ref{eq:g-ai}), which would support
random cooperation. We can further allow for repeated interaction or ``the
shadow of the future'' which are already known to foster cooperation \citep{Axelrod88,Trivers71,Binmore92}. The
important message here is that, even under worst conditions there
\emph{is} a way to reach cooperation in a game-theoretical setting by
means of social herding, i.e.~by pure social influence. Including this
additional dimension into strategic interaction avoids the lock-in into
pure defection, which is the suboptimal state compared to pure
cooperation. The mechanism we have proposed here does not rely on
additional information, in fact it uses less of the available
information, in particular no information about the payoff structure and
no comparison of alternative strategies. Further, we emphasize again the
``cost efficiency'' of the mechanism proposed in that it does not enforce
decisions agains the majority, but influences the decisions of agents
only in tie situations.

Summing up, adding social herding to strategic
interactions is a way to substantially increase the level of cooperation with \emph{less, not
  more}: simple rules instead of far-reaching regulations to enforce
cooperation, no additional information as assumed e.g.~in success driven
mechanisms, no additional costs as in other incentive schemes. Just
social herding, the right (nonlinear) way.


\bibliographystyle{ws-acs}
\bibliography{items}

\end{document}